\documentclass[conference]{IEEEtran}
\IEEEoverridecommandlockouts
\usepackage{cite}
\usepackage{amsmath,amssymb,amsfonts}
\usepackage{algorithmic}
\usepackage{graphicx}
\usepackage{textcomp}
\usepackage{xcolor}
\usepackage[normalem]{ulem}
\usepackage{multirow}
\usepackage{enumitem}
\usepackage{booktabs}
\setlist[itemize]{noitemsep, topsep=0pt, leftmargin=10pt}
\usepackage{subfigure} 
\usepackage{listings}
\usepackage{rotating} 
\usepackage[skip=1pt]{caption}
\usepackage[aboveskip=-1pt,belowskip=0pt]{subcaption}
\def\grjoin{\overset{\textsf{GR}}{\bowtie}}
\def\rgjoin{\overset{\textsf{RG}}{\bowtie}}

\newtheorem{definition}{Definition}
\newtheorem{theorem}{Theorem}

\definecolor{dkgreen}{rgb}{0,0.6,0}
\definecolor{gray}{rgb}{0.5,0.5,0.5}
\definecolor{mauve}{rgb}{0.58,0,0.82}

\newcommand{\change}[1]{#1}
\lstdefinelanguage{CypherSQL}{
    morekeywords={MATCH, WHERE, AND, RETURN, AS, SELECT, FROM, cardinality
    },
    sensitive=false, 
    morecomment=[l]{//}, 
    morecomment=[s]{/*}{*/}, 
    morestring=[b]" 
}
\lstdefinestyle{myCustomStyle}{
frame=single,
language=CypherSQL,
basicstyle=\ttfamily,
showstringspaces=false,
numbers=none,
commentstyle=\color{gray}
showstringspaces=false,
columns=flexible,
basicstyle={\scriptsize\ttfamily},
numbers=none,
numberstyle=\tiny\color{gray},
keywordstyle=\color{blue},
commentstyle=\color{dkgreen},
stringstyle=\color{mauve},
breaklines=true,
breakatwhitespace=true,
tabsize=3,
}
\lstnewenvironment{cmls}[1]{
    \lstset{
        style=myCustomStyle,
        basicstyle=\scriptsize\ttfamily, 
        abovecaptionskip=-2pt,
        belowcaptionskip=0pt
    }
}{
    \captionsetup{aboveskip=-6pt, belowskip=-1pt} 
    \captionof{lstlisting}{#1}
}

\lstnewenvironment{cmlsnc}{
    \lstset{
        style=myCustomStyle,
        basicstyle=\scriptsize\ttfamily, 
        abovecaptionskip=-2pt,
        belowcaptionskip=0pt
    }
}{}
\begin{document}

\title{MICRO: A Lightweight Middleware for Optimizing  Cross-store Cross-model Graph-Relation Joins [Technical Report]}
\author{\IEEEauthorblockN{Xiuwen Zheng}
\IEEEauthorblockA{\textit{University of California, San Diego}\\
La Jolla, USA \\
xiz675@ucsd.edu}
\and
\IEEEauthorblockN{Arun Kumar}
\IEEEauthorblockA{\textit{University of California, San Diego}\\
La Jolla, USA \\
akk018@@ucsd.edu}
\and
\IEEEauthorblockN{Amarnath Gupta}
\IEEEauthorblockA{\textit{University of California, San Diego}\\
La Jolla, USA \\
a1gupta@ucsd.edu}
}

\maketitle

\begin{abstract}
Modern data applications increasingly involve heterogeneous data managed in different models and stored across disparate database engines, often deployed as separate installs. Limited research has addressed cross-model query processing in federated environments. This paper takes a step toward bridging this gap by: (1) formally defining a class of cross-model join queries between a graph store and a relational store by proposing a unified algebra; (2) introducing one real-world benchmark and four semi-synthetic benchmarks to evaluate such queries; and (3) proposing a lightweight middleware, MICRO, for efficient query execution. At the core of MICRO is CMLero, a learning-to-rank-based query optimizer that selects efficient execution plans without requiring exact cost estimation. By avoiding the need to materialize or convert all data into a single model—which is often infeasible due to third-party data control or cost—MICRO enables native querying across heterogeneous systems. Experimental results on the benchmark workloads demonstrate that MICRO  outperforms the state-of-the-art federated relational system XDB  by up to 2.1× in total runtime across the full test set. On the 93 test queries of real-world benchmark, 14 queries achieve over 10× speedup, including 4 queries with more than 100× speedup; however, 4 queries experienced slowdowns of over 5 seconds, highlighting opportunities for future improvement of MICRO. Further comparisons show that CMLero consistently outperforms rule-based and regression-based optimizers, highlighting the advantage of learning-to-rank in complex cross-model optimization.
\end{abstract}

\section{Introduction}
In today's data-driven era, various applications in various domains, such as biomedical research, cybersecurity,  healthcare etc. \cite{moustaka2018systematic,mehta2018concurrence,lu2018internet,mageto2021big,rajendran2023patchwork,shi2023data}, are increasingly relying on heterogeneous datasets from various sources and managed across distributed sites. The outdated notion of “one size fits all” in data management no longer holds due to two major factors: 
\begin{itemize}
    \item Ownership: many datasets are controlled by third parties, making it infeasible to consolidate all into a single system; 
    \item Performance: different data models excel at different query types (e.g., graph databases for path traversal).
\end{itemize}
As a result, users often need to jointly query data stored in different models and DBMSs. To reduce the burden, \textit{polystore systems}~\cite{elmore2015demonstration,duggan2015bigdawg,gadepally2016bigdawg,alotaibi2020estocada,agrawal2016rheem} have emerged, supporting queries across heterogeneous backends.

While polystores support multiple models, query optimization remains underexplored, with no well-defined formulation across models. Key challenges include: 
\begin{itemize}
    \item the lack of a unified algebra and cost model;
    \item the autonomy of the underlying engines, which restricts access to their internal optimization logic. 
\end{itemize}  
For example, BigDAWG~\cite{elmore2015demonstration,duggan2015bigdawg,gadepally2016bigdawg}  only optimize queries within each ``island''—a single-model environment—while data movement across islands must be manually specified by users. As an initial step toward bridging this gap, this paper  studies  a specific important type of cross-model query: join queries between relational and graph data. By narrowing the scope, we aim to isolate and address key optimization challenges in this multi-model context, laying the foundation for broader research in cross-store query optimization.


\subsection{Cross-model Cross-store Workloads}
\change{Mixed queries over graph and relational data is increasingly common across applications in social sciences, cybersecurity, social media analytics, and recommendation systems. 
For instance, researchers in security analytics and data governance~\cite{houttekier2025sino,nastasa2025innovation} use citation graphs and co-authorship graphs combined with relational bibliographic and patent data to analyze anomalies in international collaborations and identifying thematic areas of undue information leakage and the parties involved.
Some prior work has studied multi-model query execution~\cite{jindal2015graph,zhao2017all,dave2016graphframes} by extending an RDBMS to support graph algorithms and optimizing such queries over an RDBMS. In contrast, we take a cross-store approach, respecting data ownership and leveraging the strengths of each native engine.}

\begin{figure}[tb!]
    \centering
\includegraphics[width=0.43\textwidth]{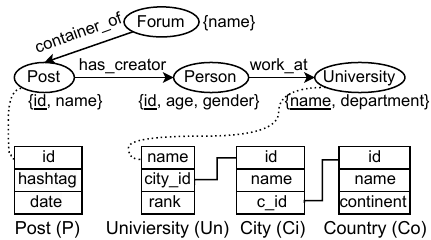}
    \caption{Example cross-models workload}\label{fig:example}
\end{figure}

\noindent\textbf{\textit{Motivating Workload.}} We consider a real-world \change{cross-store} cross-model query scenario. As shown in Fig.~\ref{fig:example}, the graph is derived from a benchmark such as LDBC~\cite{erling2015ldbc}, where nodes represent entities (forums, posts, persons, universities), and edges, their relationships. Relations come from external datasets such as forum platforms, QS university rankings, and Wikipedia: \texttt{Post} (\texttt{P}) with post metadata, \texttt{University} (\texttt{Un}) with city and ranking info, \texttt{City} (\texttt{Ci}) mapping cities to countries, and \texttt{Country} (\texttt{Co}) mapping countries to continents. A representative query retrieves hashtags of posts authored by people in \textit{top-ranked} \textit{Computer Science} departments at \textit{Asian} universities, along with the corresponding forums.

A naive plan is to run the graph query, then move results to the relational engine for joins: a path traversal with a filter on \texttt{University} node property department retrieves all \texttt{(Forum, Post, University)} tuples (forum name, post id, university name). The results are then joined with tables using rank and continent filters. However, this plan can be inefficient: dense graphs yield large intermediate results, incurring high transfer costs.

Many alternative plans exist. For example, table \texttt{P} can be materialized as  nodes (e.g., labeled \texttt{PR}) in the graph, enabling a node join with existing  \texttt{Post} nodes. If  \texttt{P} is small and \texttt{Post} nodes are central (high-degree), this plan can greatly reduce  graph query result size and  the subsequent data movement cost.   
However, if \texttt{P} is  large, moving it to the graph is costly and   the node join may introduce significant overhead. If the node joins are not selective, it may not reduce the graph query size much.  Thus, determining which relations to move across systems is key to optimizing cross-model query execution.

\subsection{Challenges} \label{sec:chanllenge}
Optimizing this kind of cross-models join queries to find the best plan out of many candidates can be very challenging. We elaborate four main challenges as below:

\textit{\textbf{C1: Development of a unified cost model.}}  Each underlying DBMS has its own cost model, e.g., PostgreSQL's cost model yields relative cost units, while Neo4j's generates estimated cardinalities, making cross-engine cost comparisons infeasible. Additionally, these models  depend on internal statistics (e.g., histograms, value distributions) to estimate cost. When sub-queries involve foreign data which is intermediate results produced by other systems and transferred at runtime, these statistics are missing, making cost estimation unreliable.


\textit{\textbf{C2: Adaptiveness to dynamic environments.}} Federated databases are often deployed in distributed settings with varying hardware, locality, and network conditions. The cost of data movement across systems can differ significantly based on whether databases are co-located or geo-distributed. A practical middleware must be adaptive to these environmental factors to choose efficient plans.

\textit{\textbf{C3: Lack of expressiveness in existing algebra.}} While prior work~\cite{marton2017formalising,thakkar2017towards} has extended relational algebra to support graph queries, existing frameworks still lack the expressiveness to represent cross-model queries. Similar to single-model optimizers, a unified algebraic foundation is needed to enable plan enumeration and transformation. 

\textit{\textbf{C4: Enlarged planning space.}} Unifying relational, graph, and cross-model operations into a shared algebra significantly increases the query plan search space. The optimizer must consider a combinatorial number of join orders and different data movement strategies. Even if an optimal plan is selected, the underlying DBMSs may not execute delegated sub-queries exactly as intended, introducing additional uncertainty in the actual runtime behavior.

\subsection{Solution To The Challenges} \label{sec:intro-lero}

To address  these challenges, we  build upon decades of research in RDBMSs, adapting their insights and techniques to this new context.

Recent work has explored using Machine Learning (ML)  to help build query optimizers for both relational DBMSs and federated systems~\cite{sun2019end,zhi2021efficient,hilprecht2022zero,marcus2021bao}. Many of these approaches propose a \textit{learned cost model} or plan value function to estimate the quality (e.g., latency) of each candidate plan to replace the traditional cost models. However, training such  models is often challenging: it  requires a large number of training data  considering the model complexity.  In our cross-model, multi-store setting, the problem is even more complex since the total cost consists of multiple parts and involves two data models.  

A  recent system, Lero~\cite{zhu2023lero}, states that predicting  exact latency for each plan is an overkill since the goal of optimizer is to find the best execution plan. Instead of pointwise prediction (assigning a numeric cost to each plan), Lero proposes a learning-to-rank framework using a pairwise comparison model: given two plans, the model learns to predict which one is better. Their experiment proves  that comparing to a pointwise regression  model, training a binary classifier requires much fewer training samples.

\textit{\textbf{S1: Predicting exact costs can be avoided.}} We adapt the  learning-to-rank paradigm.  C1 can be  addressed by avoiding the need to predict exact cost.

\textit{\textbf{S2: Learned model has better generalization.}}  Unlike traditional rule-based or heuristics-based optimizers that rely on magic constants, learned models uses model parameters and can adapt to 
new system environments as long as sufficient training data is available.

\textit{\textbf{S3: Introduction of cross-model join operators.}} Previous work~\cite{marton2017formalising,thakkar2017towards} has
extended relational algebra to support graph queries by proposing  graph relation and define operators on it. We define two new operators for joining between graph relations and relational tables. These operators enable algebraic expression  for cross-model queries.

\textit{\textbf{S4: Lightweight middleware architecture.}} We  build a minimal middleware  that focuses only on high-level decisions  keeping the planning overhead low while enabling effective performance gain. It preserves the autonomy of  individual engines, does not require modifications to their internal code which makes deployment easier, and delegates most optimization tasks to their native optimizers. 

\subsection{Contributions}
We present \textbf{MICRO}, a lightweight \underline{MI}ddleware for \underline{CRO}ss-model cross-engine queries. At its core  is \textbf{CMLero} - a \underline{C}ross-\underline{M}odel \underline{Le}arning-to-\underline{r}ank query \underline{o}ptimizer  that selects  execution plans with low  latency. Key contributions  are:

\begin{itemize}
    \item We formally  define a class of  cross-model join queries between graph and relational data, where a graph traversal is joined with  relational tables via vertex properties and table columns. We provide a uniform algebra to express them.
    \item To the best of our knowledge, MICRO is the first to  optimize such joins  in a federated, cross-engine  setting.
    \item We introduce a learning-to-rank framework for query optimization in the cross-model setting.  It  avoids the overhead of  estimating exact  costs of plans  which  is  a complex task  under cross-models cross-engines settings.
     \item We publish one real-world and four semi-synthetic benchmarks combining graph and relational data with hundreds of queries, benefiting future research in this area.

    \item MICRO is a lightweight middleware built  on  existing databases. It  leverages     mature native query optimizers while preserving   autonomy, requiring no modifications to internal code and  ensuring ease of adoption and extensibility.

\end{itemize}
\noindent The benchmark and MICRO code are publicly available to support reproducibility and future research.\footnote{https://github.com/xiz675/MICRO}




\section{System Architecture} \label{sec:sys}
\begin{figure}[tb!]
    \centering
    \includegraphics[width=0.5\textwidth]{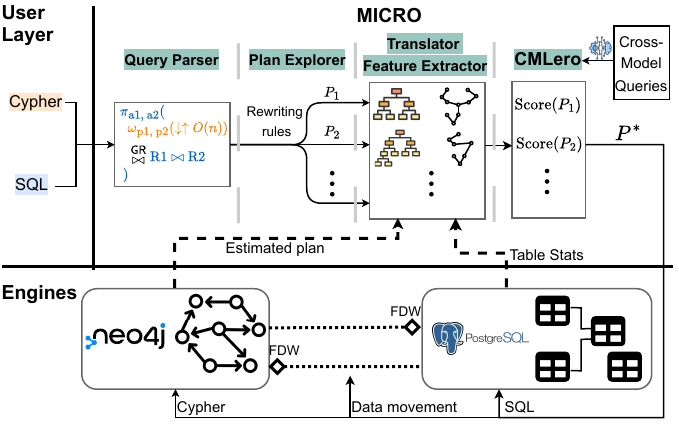}
    \caption{System architecture for MICRO}\label{fig:sys}
\end{figure}

Fig.~\ref{fig:sys} provides an overview of  MICRO architecture. MICRO is built on top of a graph database and a relational database, and takes as input a graph and a relational query. The system first parses both queries into a unified algebraic expression. A plan explorer then enumerates all candidate execution plans by rewriting the algebraic expression into equivalent forms. Each candidate plan is translated into corresponding graph and relational subqueries, along with the  data movement operations for cross-model joins. Then they are encoded as feature vectors using statistics  collected from the underlying databases. Then the learned comparator model, CMLero, will rank the plans. The top-ranked plan is selected and executed by delegating subqueries to the respective databases and transferring intermediate results across engines.

\noindent\underline{\textit{Extensibility of the system.}} The figure uses  Cypher and SQL as example query languages for  property graph and relations.  MICRO is not tied to specific languages or engines. It can support other standard query languages—e.g., Gremlin for graphs—as long as appropriate parsers and translators between query syntax and the unified algebra are provided. Similarly, while PostgreSQL and Neo4j are used in our prototype, MICRO can integrate other engines, provided that the necessary features required by the learned comparator can be extracted.

\noindent\underline{\textit{Organization}}. The rest of the paper is organized as follows. Section~\ref{sec:preliminary} introduces background on relational graph algebra. Section~\ref{sec:algebra} presents our unified algebra for cross-model queries and formally defines the class of queries addressed in this work. Sections~\ref{sec:design}  formalizes the  query optimization problem and describes the plan exploration framework. Section~\ref{sec:model} introduces the learned comparator model, CMLero. Section~\ref{sec:benchmark} describes the benchmarks. Section~\ref{sec:experiment} presents our experimental results. Finally, Sections~\ref{sec:related} and \ref{sec:con} discuss related work, conclude the paper, and outline future directions.


\section{Preliminary}\label{sec:preliminary}
We briefly introduce the \textit{Property Graph} data model, which is widely adopted by modern graph databases, and review prior work on extending relational algebra to support this model.

\subsection{Data Model}
We adapt the definition of property graph in  \cite{angles2018property,marton2017formalising,thakkar2017towards}.
\begin{definition}[Property Graph] 
Let $L_v, L_e$ be a set of vertices and  edge labels respectively; $P_v, P_e$ be a set of property names for vertices and edges respectively; $D$ denote the domain  of atomic values of all properties; $\text{SET}^+(X)$ be all subsets of set $X$.  A property graph  is defined as $$G= \{V, E, \rho, \lambda_v, \lambda_e, \mu_v, \mu_e\},$$
where 
\begin{itemize}
    \item $V$ is a set of vertices; 
    \item $E$ is a set of edges;
    \item $\rho: E\to V \times V$ is a function that associates each edge in $E$ to a pair of source and target vertices. 
    \item $\lambda_v:V\to\text{SET}^+(L_v)$ (or $\lambda_e: E \to \text{SET}^+(L_e)$) assigns vertex (or edge) labels  to  vertices in $V$ (or edges in $E$).
    \item $\mu_v: V\times P_v \to D \cup \{\epsilon\}$ (or $\mu_e: E\times P_e \to D$) assigns each vertex (or edge) property to a value in $D$ or a NULL value $\epsilon$ if the vertex (or edge) does not have that property. 
\end{itemize}
\end{definition}
\noindent To extend relational algebra to the property graph model, \cite{marton2017formalising} introduced the notion of a \textit{graph relation}, which allows graph elements to be represented relationally.

\begin{definition}[Graph Relation]
    Given a property graph $G= \{V, E, \rho, \lambda_v, \lambda_e, \mu_v, \mu_e\}$, a \textit{graph relation} $r$ is a bag of tuples so that $\forall{A} \in sch(r) : dom(A) \subseteq V \cup E \cup D,$
where  $sch(r)$ is its schema  and $dom(A)$ is the domain of attribute $A$.
\end{definition}


\subsection{Relational Graph Algebra}
We  first recap some  \textit{relational algebra operators} in Table~\ref{tab:reope}. Prior work \cite{marton2017formalising,thakkar2017towards} has proposed \textit{relational graph algebra}, which  extends relational algebra with graph-specific operators  for  OpenCypher and Gremlin queries. In this work, we adapt and refine a subset of the operators proposed in \cite{marton2017formalising}, and present them in Table~\ref{tab:ope}. 
To avoid confusion, we use distinct notations to differentiate graph-specific operators from their relational counterparts. In schema definitions, we use $||$ to indicate column concatenation (i.e., appending attributes), and $\setminus$ to denote attribute removal.

\begin{table}[b]
\caption{Relational algebra operators} \label{tab:reope}
\centering
\begin{tabular}{|l|c|p{4cm}|}
\hline
\textbf{Operation} & \textbf{Notation} & \textbf{Description} \\ \hline
Join               & \change{$r_1 \bowtie r_2$}       & Joins two relations \change{$r_1$} and \change{$r_2$} on their common attributes. \\ \hline
Projection         & \change{$\pi_{a_1, a_2,\ldots}(r)$} & Extracts specified attributes (columns) from a relation \change{$r$}. \\ \hline
Selection          & \change{$\sigma_{\text{condition}}(r)$} & Filters tuples from relation \change{$r$} that satisfy a given condition. \\ \hline
\end{tabular}
\end{table}


\begin{table}[bt]
\centering
\setlength{\tabcolsep}{1pt}
\caption{Relational graph operators}
\label{tab:ope}
\small
\begin{tabular}{|c|c|c|c|}
\hline
&\textbf{Operation} & \textbf{Notation}  &  \textbf{Schema}  \\ \hline
\multirow{3}{*}{\begin{tabular}[c]{@{}l@{}}Relation\\ Analogous\end{tabular}}  &Selection & $\gamma_{\text{condition}}(p)$ & $\text{sch}(p)$\\\cline{2-4}
    &Projection & $\omega_{x_1, x_2, \cdots}(r)$ & $\langle x_1, x_2, \cdots\rangle$\\ \cline{2-4}
& Join & $p_1 \otimes p_2$ & 
$\text{sch}(p_1) \| (\text{sch}(p_2)\setminus\text{sch}(p_1))$
\\ \hline
\multirow{2}{*}{\begin{tabular}[c]{@{}l@{}}Graph\\ Extension\end{tabular}}
&Get-vertices   & $O(v:L)$  & $\langle v\rangle$\\ \cline{2-4}
&Expansion &$\updownarrow_{v_1}^{v_2:L}[:T](p)$ & $\text{sch}(p) \| \langle v_2\rangle$ \\
& &\change{$\updownarrow_{v_1}^{v_2:L^+}[:T](p)$} & \change{$\text{sch}(p) \| \langle v_2\rangle$} \\
\hline 
\end{tabular} 
\end{table}

\underline{\textit{Basic Operators.}}
Some graph operators have direct analogues in relational algebra.   Selection operator $\gamma_{\text{condition}}(r)$ filters tuples from a graph relation that satisfy a given condition. Projection operator $\omega_{x_1, x_2, \cdots}(r)$ extracts certain  attributes  from the graph relation. Join operator $\otimes$ is analogues to relational join  $\bowtie$: it performs a Cartesian product between two graph relations and filters tuples based on common attribute values. 

\underline{\textit{Graph-specific Extensions.}} Other operators are unique to the graph context.  The get-vertices operator $O(v:L)$ retrieves all vertices with label $L$, returning a relation with a single attribute $v$. The path expansion operators $\uparrow_{v_1}^{v_2:L}[:T](p)$ and $\downarrow_{v_1}^{v_2:L}[e:T](p)$  traverse from a vertex $v_1$ to an  adjacent vertex $v_2$ with label $L$ in the outgoing or incoming direction
via an edge with label $T$, \change{and $\updownarrow_{v_1}^{v_2:L^+}[:T](p)$ traverse from  $v_1$ to $v_2$ by one or more $T$ edges.} These operators extend the schema of the input graph relation  with two additional attributes representing  vertex $v_2$.

\underline{\textit{Example.}} With these operators, the graph pattern matching query from our  motivated example can be expressed as:
$$
  p = \uparrow_{n_2}^{(n_3:\text{University})}\uparrow_{n_1}^{(n_2:\text{Person})}\uparrow_{n_4}^{(n_1:\text{Post})} O(n_4:\text{Forum}),  
$$
where edge types and selection operators are omitted for brevity.
To extract needed information, apply a project operator: $
X = \omega_{n_1.\text{id} \to \text{pid}, n_4.\text{name} \to \text{fname}, n_3.\text{name} \to \text{uname}}(p)$
\section{\change{Cross-Model Query Definition}} \label{sec:algebra}
To support unified cross-model query processing, we extend relational algebra with operators that enable joins between graph and relational data. This integrated algebra forms the foundation for formally defining cross-model join queries.

\subsection{Operators for Cross-Model Joins}
We introduce two operators to support joins between relational tables and graph vertices, summarized in Table~\ref{tab:cmope}. Each operator transforms one data model into the other and applies a native join in the destination model.

\underline{\textit{Relation-Graph Join.}} The operator $r \rgjoin p$ first transforms the relational table $r$ into a graph representation by mapping each tuple to a vertex and its attributes to vertex properties. A temporary label $l \notin L_v$ is assigned to these new vertices to avoid conflicts with existing graph labels. These vertices are then joined with the graph relation $p$ using a graph-native join operator on vertex properties.
Formally: $r\rgjoin p = \omega_{a_1, a_2, \cdots}(O(n:l)) \otimes p$ where ${a_i}$ are properties used to match with vertices in $p$. The output is  a graph relation. 
 
\begin{table}[bt]
\centering
\caption{Cross-model Join Operators}
\label{tab:cmope}
\setlength{\tabcolsep}{1pt}
\small
\begin{tabular}{|c|c|c|}
\hline
\textbf{Operation}   & \textbf{Notation}   & \textbf{Output and Schema}     \\ \hline
Relation-Graph Join & $r\rgjoin p$  &  graph relation, $\text{sch(r)}\|(\text{sch(r)}\setminus\text{sch(p)})$  \\ \hline
Graph-Relation Join & $p \grjoin r$ &  relation, $\text{sch(p)}\|(\text{sch(p)}\setminus\text{sch(r)})$  
\\\hline 
\end{tabular} 
\end{table}


\underline{\textit{Graph-Relation Join.}} The operator  $p \grjoin r$  first materializes the graph relation $p$ as a relational table $r_p$, then joins it with the table $r$ using relational join operator.  Formally: $p \grjoin r = r_p \bowtie r$. The output is a relational table. 


With these integrated operators, the motivated workload in Fig.~\ref{fig:example} can be expressed as:
\begin{subequations} \label{eq:example}
\begin{align}
&\pi_{\text{fname, hashtag}}( \label{eq:1a}\\  
&\ \ \ \ \ \ \ \ X \grjoin_{\text{pid=P.id}} \text{P} \bowtie_{\text{uname=Un.name}} \text{Un} \bowtie \text{Ci} \bowtie \text{Co}), \label{eq:1b}  
\end{align}
\end{subequations}
Equation~\ref{eq:1b}  applies  graph-relation join to materialize the graph relation $X$ as a relational table to  join with   the other tables by relational joins.  \ref{eq:1a} projects the output attributes  using the  relational projection operator.








\subsection{Cross-model Join Query}
Using the relational graph operators introduced, we formalize the notion of a graph query as follows:
\begin{definition}[Graph query]
A graph query can be expressed as an  algebraic expression  using the relational graph operators listed in Table~\ref{tab:ope}.
\end{definition}

Building on this, we define a specific class of cross-model join queries that are the focus of this paper:
\begin{definition}[Cross-model graph-relation join (CMGRJ) query]
A CMGRJ query consists of a \textit{single}   graph query;  and a subsequent \textit{single} relational query over the  graph query result materialized as a relation and relations, expressed using the  relational operators  in Table.~\ref{tab:reope}. 
\end{definition}

More generally, \change{modern graph and relational databases support more operators (e.g., system-specific analytical ones). However, for the purposes of designing the  benchmark queries and presenting a generalizable architecture, we focus on a core set of basic operators.}
There are different patterns of cross-model join queries, such as those where a complex relational query followed by a graph query that joins the materialized vertices of the relational query result with vertices of one  or multiple graphs. The optimization of this type of queries is analogous to that of the CMGRJ queries. This paper mainly focuses on CMGRJ queries. 


\section{Design Decision, Optimization Problem and Plan Explorer}\label{sec:design}
\begin{figure*}[ht!]
    \centering
    \includegraphics[width=0.98\textwidth]{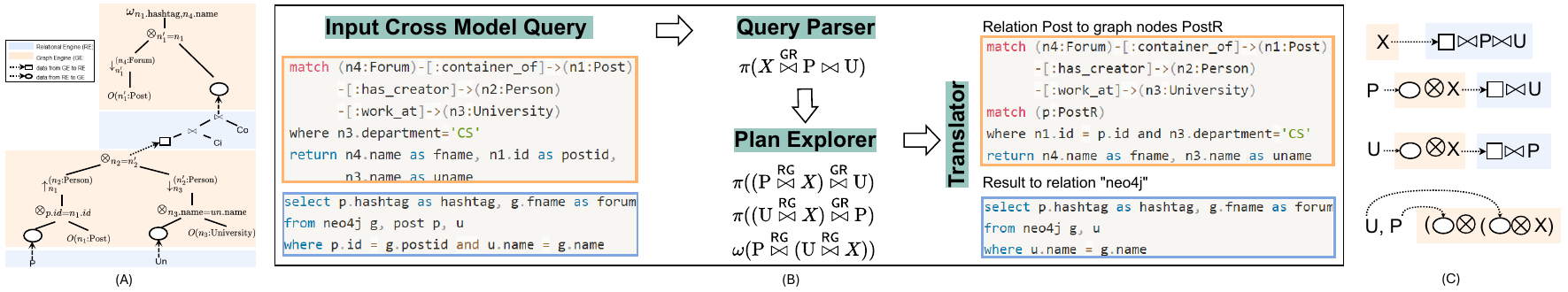}
    \caption{(A) An example plan for motivated workload (B) Plan explorer framework; (C)  Candidate plans for motivated workload.}\label{fig:planexplore}
\end{figure*}

We revisit challenge \textbf{C4} from Section~\ref{sec:chanllenge} through a concrete example. 
Given a CMGRJ query, the combination of relational, graph-specific, and cross-model operations introduces a vast plan space.
For the motivated workload, Fig.~\ref{fig:planexplore} (A) shows a possible plan, where the blue-shaded regions denote  operations performed in the relational database, while yellow-shaded regions denote operations in the graph database.

In this plan, the tables \texttt{P} and \texttt{Un} are  moved to the graph database,  transformed into vertices and joined with the \texttt{Post} and \texttt{University} vertices on properties, \texttt{id} and \texttt{name}. The  partial path from \texttt{Post} $\to$ \texttt{Person} $\to$ \texttt{University} is then materialized as  a relation  and  joined with \texttt{Ci} and \texttt{Co} in the relational database. The resulting table is moved back to the graph database and  joined with another partial  path \texttt{Forum} $\to$ \texttt{Post} on shared \texttt{Post} vertices.

Many alternative plans are possible by reordering joins, traversal steps, or choosing different points for data transfer across systems. The joint optimization of relational, graph, and cross-model operations results in an exponentially growing plan space. Even within the graph engine, traversal orders affect performance; similarly, the relational engine offers a large join-order space. Adding cross-model joins further increases complexity due to decisions on  where each join should occur and  their associated data movement costs.

\subsection{Problem Definition}
To address this complexity, we design a lightweight middleware that avoids exhaustively searching the entire plan space. Instead, it focuses on high-level strategic decisions about data movement and join execution locations \change{for  cross-model vertex-relation joins}. \change{It does not decompose the graph query; rather,} the middleware operates as follows:

\begin{itemize}[leftmargin=10pt]
    \item Step 1: Execute filters and joins over  relational data in  relational engine and materialize the results as tables; \change{apply filters on node properties and assign these nodes a new label.}
    \item \change{Step 2: Identify vertex types to move to relational engine, join with corresponding tables with results stored as tables.}
    \item Step 3: Identify selected tables \change{and the join results} to transfer to the graph engine  and materialize them as vertices;
    \item Step 4: Perform joins between new vertices and the original graph query in the graph engine;
    \item Step 5: Transfer the graph query result back to the relational engine and complete the remaining joins there.
\end{itemize}
\change{In Step 3, vertex–table join results must be moved back to the graph engine to participate in the path traversal graph query.}
Under this design, we formalize the  optimization problem:
\begin{definition} [Cross-store Cross-model query optimization]
Given a property graph in a graph database  and a set of relations  in a relational database, for a CMGRJ query, select \change{vertices (by label)  to materialize as tables in the relational database and }  relations  to  materialize as vertices  in the graph database. 
The objective is to minimize  end-to-end latency which includes: 1) data transfer cost across databases, and 2)  native query execution time within each database.  
\end{definition}

\subsection{Plan Explorer} \label{sec:planexp}
Fig.~\ref{fig:planexplore} (A) illustrates the plan generation process for the motivating workload. A CMGRJ query, consisting of a Cypher query and a SQL query, is first parsed into an integrated algebraic expression. The relational filters and joins are pre-computed and materialized as relation \texttt{U}:  $\sigma_{rank<500}(\text{Un})\bowtie \text{Ci} \bowtie \sigma_{continent='Asia'}(\text{Co})$. The original algebraic expression derived from the query is simplified as: $\pi_{\text{fname, hashtag}}(X \grjoin_{\text{pid=P.id}} \text{P} \bowtie_{\text{uname=U.name}} \text{U})$ which is the raw plan \change{with no data movement}.  Candidate plans are generated by \change{choosing different vertices and tables to materialize in the other engine.} Each plan is translated  to new  Cypher and SQL queries with explicit data movement instructions. We omit parser and translator  details and focus on plan generation.


\underline{\textit{Plan Space Complexity.}}  
\change{For each vertex–relation join, there are three options: (1) move vertices to the relational engine, (2) move the relation to the graph engine, or (3) keep both in place and join the graph query result with the relation in the relational engine.} We state a theorem on the  plan space size.
\begin{theorem} 
Let $n$ be the number of relational tables with join attributes matching graph vertex properties, \change{then the number of candidate plans is $O(3^n)$}. 
\end{theorem}



\change{\subsection{Pruning Rules}}
\change{One possible plan for the motivating query is to move the Post vertices to the relational engine, joined with table \texttt{P}, and then transferred back to the graph engine  to participate in the graph query. The plan is expressed by $\pi_{\text{fname, hashtag}}((O(:\text{Post}) \grjoin \text{P})\rgjoin X \grjoin  \text{U})$.}

\change{Alternatively,  table \texttt{P} can be moved to the graph engine:  
$\pi_{\text{fname, hashtag}}(\text{P} \rgjoin X \grjoin \text{U})$.  
The key difference is that the first plan materializes the post entity join in relational engine before executing the graph query, while the second lets the graph engine optimize the entity join with the original graph query natively by moving Post table to graph engine. The first plan also incurs two data transfers from graph  to relational  engine (two $\grjoin$ operators) which can be costly.}

\change{In general, plans that move vertices to the relational engine add additional data transfers and force materialization for certain entity joins, limiting the benefit of the native graph optimizer. We therefore prune such plans. In the 5-step framework, we remove Step 2 and only consider table movements. This reduces the plan space from $3^n$ to $2^n$. In the motivated example, the plans drop from 9 to 4.  Fig.~\ref{fig:planexplore} (C) shows these plans where circles denote relations materialized as graph vertices, and rectangles denote intermediate results transferred to the relational engine. This pruning strategy significantly simplifies planning while preserving good performance.}

\subsection{Lightweight Middleware}
We justify the design of our plan explorer and pruning rules along four dimensions.

\underline{\textit{Reducing Planning Complexity.}} By delegating intra-model planning to native engines, the middleware avoids enumerating the full cross-product of possible plans across models, reducing complexity and overhead.

\underline{\textit{Leveraging Native Optimizers.}} Graph systems excel at traversal and pattern matching, while relational systems excel at joins and filtering. The middleware reuses these native optimizers rather than re-implementing them.

\underline{\textit{Preserving  Autonomy and Generalizability.}} Our design does not impose detailed plans on underlying engines. This preserves engine autonomy, maintains compatibility with future updates, and enables extension to other systems.

\underline{\textit{Enhancing Scalability and Reducing Data Movement.}}
Data movement is often the primary bottleneck in cross-system queries~\cite{gavriilidis2023situ}. Data transfers are limited to at most one in each direction (relational $\rightarrow$ graph and graph $\rightarrow$ relational), avoiding multiple round-trips,  unlike the plan in Fig.~\ref{fig:planexplore} (A) (two relational $\rightarrow$ graph transfers) or pruned plans with two graph $\rightarrow$ relational transfers. Filters and joins are also pushed down to the relational engine to shrink the data moved initially.

\section{Learned Comparator Model} \label{sec:model}
The plan comparator CMLero is a learned model designed to rank all the plans generated by the plan explorer  for a given CMGRJ query. The structure of CMLero is shown in Fig.~\ref{fig:model}.

\subsection{Feature Extractor}
As shown in Fig.~\ref{fig:planexplore} (A), a candidate plan from the translator includes data movement, a modified Cypher and SQL query. We extract features from each plan. 

\subsubsection{Plan structure}
\begin{figure*}[ht!]
    \centering
    \includegraphics[width=0.99\textwidth]{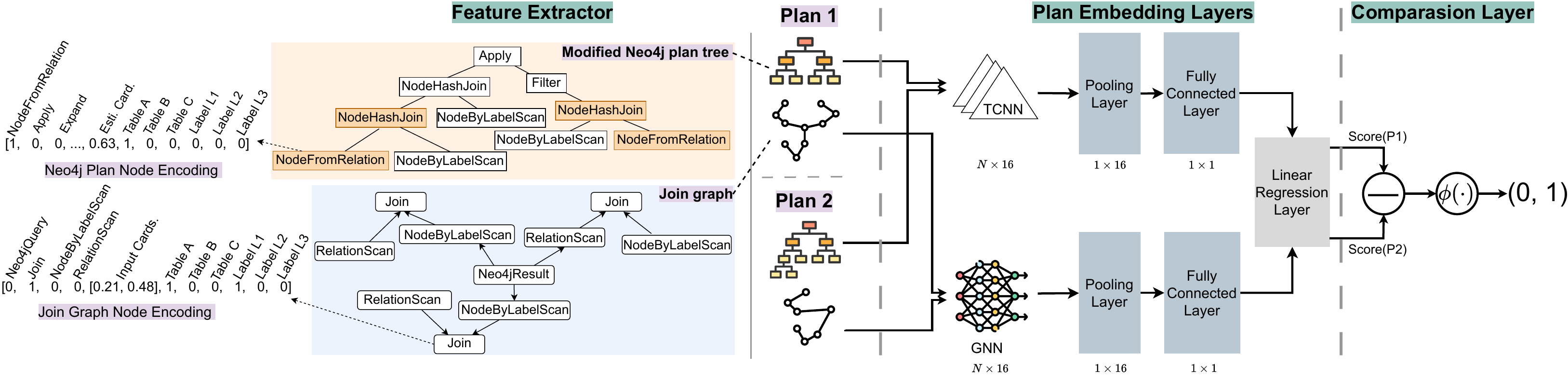}
    \caption{CMLero: Learned comparator model structure.}\label{fig:model}
\end{figure*}

As Fig.~\ref{fig:model} feature extractor part shows, each   plan  consists of two parts:

\textit{\underline{Modified Neo4j Plan Tree.}} A tree-structured representation of the raw Cypher query plan, modified to account for data movement. If the Cypher query includes external tables moved from the relational DBMS, Neo4j cannot generate a complete plan. Thus, we first generate a query plan tree for the original Cypher query without external data, then modify it to include new nodes: \texttt{NodeFromRelation}: Represents tables moved to Neo4j and transformed into vertices; and \texttt{NodeHashJoin}: Represents joins between the new nodes and existing graph nodes on common node properties. These modifications integrate external data into the Neo4j plan tree structure.
When a table \texttt{A} is moved and joined with label \texttt{L} nodes,  the original plan tree will be traversed to find the plan node, say $N$, that firstly visits label \texttt{L} nodes.  Then a \texttt{NodeHashJoin} plan node will be inserted so that its parents inherits the parents of $N$ and  $N$ becomes its child node, and a \texttt{NodeFromRelation} plan node will be added to be the other child of \texttt{NodeHashJoin}. This modified plan may not be the same as the real plan  executed by Neo4j after the tables are actually moved there, but it captures the cross-model joins information.

\textit{\underline{Join graph.}} A graph-structured representation of the SQL query, highlighting the star join pattern between the materialized graph relation and other relational tables since all the relational joins and filters on relations only are pre-computed and materialized. Operators include 1) \texttt{Neo4jResult}: Materializes the Cypher query result as a relation; \texttt{NodeByLabelScan} and \texttt{TableScan}: Project relevant columns from the graph relation or native tables; and \texttt{Join}: Represents relational joins.

\subsubsection{Plan Node Encoding}
Each plan node in the modified Neo4j plan tree or join graph is encoded into a feature vector. These vectors capture:

\underline{\textit{Node type.}} One-hot encoded for operator type.

\underline{\textit{Cardinality.}} In the Neo4j plan tree, Neo4j provides estimated output  cardinality for  native plan nodes. For the newly inserted  nodes: we assign output  cardinality of \texttt{NodeFromRelation} as  the  relation size, and that of \texttt{NodeHashJoin} as  unknown. In the join graph, we instead use input cardinality for each operator node.   For example, \texttt{Join} node has two children: \texttt{RelationScan} and \texttt{NodeByLabelScan}, thus  its input cardinalities  will be the  relation size and the estimated size of the graph query result. Like Lero, a min-max normalization is applied to  the logarithmic  cardinality values to account for their wide range.

\underline{\textit{Touched tables and labels.}} Binary indicators for all tables and vertices labels that have been visited by the current  node and all its descendants. For example, the \texttt{Neo4jResult} node in the join graph contains all the tables and labels that are touched by the modified Neo4j plan tree.

\subsection{Model Structure}
Similar to the Lero model \cite{zhu2023lero}, CMLero comprises of  plan embedding layers and a comparison layer. It takes two encoded plans,  maps each to a one-dimensional value through its embedding layers, and compares the two values to output a score $p \in [0, 1]$ that indicates which plan has lower latency.

\subsubsection{Plan Embedding Layers}
After encoding each plan node, the modified Neo4j plan tree is a tree structure of vectors and the join graph is a graph of vectors. Unlike Lero, which only embeds a tree structure, CMLero embeds both a tree and a graph structure. The embedding layers consist of:

\underline{\textit{Tree embedding layers.}}
Tree embedding layers use Tree-Based Convolution Neural Network (TBCNN)~\cite{zhu2023lero,10.14778/3342263.3342644,mou2016convolutional}. 
Different from a general CNN which uses a small matrix of weights to slide across an input image, TBCNN uses a binary tree kernel (a parent node and two children node) to slide over an input tree. After applying three  TBCNN layers, the original tree is transformed into a new tree with dimension $N\times 16$ ($N$ is the number of plan nodes in the original tree) where each node aggregates the information of itself and its two children. Then a dynamic pooling layer aggregates the   nodes features to generate a tree embedding with dimension $1\times 16$. Lastly, a fully connected layers maps the aggregated tree features into a scalar latency-related score.

\underline{{\textit{Graph embedding layers.}}}  Graph Embedding consists of two-layer Graph Convolution Network (GCN) that update each  node's feature by aggregating its neighbors' features, a dynamic pooling layer that  summarizes the entire graph into a fixed-length vector which is 16, and a fully connected layer to reduce this to a single latency-related scalar.

\underline{\textit{Linear regression layer}} computes the weighted sum of the two scalar output from tree and graph embeddings to  produce a final scalar latency-related score  of  the plan.

\subsubsection{Plan Comparator}
Let $\mathcal{P}(q)$ be the candidate plans set for a query $q$,   given two candidate plans $P_i, P_j \in \mathcal{P}(q)$, let $\text{Emb}(P_i), \text{Emb}(P_j)$ be their scalar embeddings. A sigmoid function is applied on their difference:
$$\hat{y_{ij}} = \sigma(\text{Emb}(P_j) - \text{Emb}(P_i)) = \frac{1}{1+e^{-(\text{Emb}(P_j) - \text{Emb}(P_i))}}$$
$\hat{y_{ij}} \in [0, 1]$ indicates the probability that $P_i$ is better than $P_j$ in terms of latency. $\hat{y_{ij}} > t$ where $t$ is a threshold value indicates $P_i$ is preferable, a higher $\hat{y_{ij}}$  means $P_i$ is more preferable; when $\hat{y_{ij}} \leq t$, $P_j$ is preferable. 

\subsubsection{Loss Function}
Let  $y_{ij}\in \{0, 1\}$ be the true label. Let $y_{ij}=1$ if  $P_i$ has lower latency  and $0$ if $P_j$ is better with lower latency, then the likelihood of $y_{ij}$ subjects to a Bernoulli distribution: $P(y_{ij}|\hat{y_{ij}}) = \hat{y_{ij}}^{y_{ij}}\cdot(1-\hat{y_{ij}})^{(1-y_{ij})}$.
The log-likelihood would be:
$$\log P(y_{ij}|\hat{y_{ij}}) = y_{ij} \log(\hat{y_{ij}}) + (1-y_{ij})\log(1-\hat{y_{ij}})$$

For  a set of workload queries $\mathcal{Q}$,   the goal is to maximize  likelihood over  all candidate plan pairs of  all queries,  equivalent to minimizing the negative log-likelihood over all samples:
\begin{equation}\label{eq:likelihood}
\min L = -\frac{1}{N} \sum_{q\in\mathcal{Q}}\ \sum_{1 \leq i < |\mathcal{P}(q)|}\ \sum_{i < j \leq |\mathcal{P}(q)|} \log P(y_{ij}|\hat{y_{ij}})    
\end{equation}
where $N$ is the total size of samples.

\subsection{Model Training}
The embedding and comparison layers are shared for two plans, so  only one model needs to be trained. To train  CMLero, i.e., updating  parameters in the plan embedding layers, we use stochastic gradient descent (SGD) on a large set of  training samples.  Each sample consists of a pair of plans 
with a binary label indicating which plan has lower latency.

For each  training query $q \in \mathcal{Q}$, the plan explorer generates all the candidates $\mathcal{P}(q)$. These plans are   executed to collect the actual latencies. For each  pair  $(P_i, P_j)$, a label is assigned based on  which  has lower latency (1 if $P_i$ is better, 0 otherwise), yielding   $N_q = \frac{|\mathcal{P}(q)|(|\mathcal{P}(q)|-1)}{2}$ training samples. The total number of training samples over $\mathcal{Q}$ is: $N = \sum_{q\in \mathcal{Q}}N_q$. Model parameters are trained to minimize Equation~\ref{eq:likelihood}. SGD enables efficient optimization across this large dataset.
\subsection{Inference}
Once trained, CMLero ranks candidate plans for a test query $q_t$. The plan explorer generates $\mathcal{P}(q_t)$. Each plan is represented by a Neo4j plan tree and a join graph with each node encoded by feature vector. These are passed through the embedding layers to produce a single ranking score. Plans are then sorted by score, with lower scores indicating lower expected latency. The top-ranked plan is selected for execution.

\section{Benchmark}\label{sec:benchmark}
To the best of our knowledge, there are no publicly available benchmarks for evaluating cross-database, cross-model query systems. To fill the gap and assess the performance of \textsc{MICRO}, we develop two benchmarks: the \textbf{OpenAlex-USPTO Benchmark}, built from real-world heterogeneous data, and the \textbf{CM-LDBC Benchmark}, a semi-synthetic dataset extending the LDBC graph benchmark~\cite{erling2015ldbc} with relational components. Both are publicly released to support future research in cross-model query optimization.

\subsection{OpenAlex-USPTO (OA-USPTO) Benchmark}
\subsubsection{Dataset.}
OpenAlex~\cite{priem2022openalex} is an open catalog of global scholarly research. We extract millions of recent publications and associated metadata—authors, institutions, and topics—filtered by: (1) physical sciences domain, (2) over 10 citations, and (3) published after 2010. From this, we construct a graph dataset summarized in Tables~\ref{tab:openalexnode} and~\ref{tab:openalexedge}.

We also collect recent patents (filed after 2022) from the USPTO Open Data Portal~\cite{uspto} that mention keywords like ``large language model,'' ``artificial intelligence,'' and ``machine learning,'' yielding 124,110 patents. Metadata includes inventors, affiliated institutions, and cited publications are also collected as relations. The relational schema and statistics are shown in Table~\ref{tab:usptodata}. Fields in \textbf{bold} denote derived or preprocessed attributes discussed in the next subsection.

\subsubsection{Entity Matching}
To enable cross-model joins, we resolve entities across the graph and relational datasets referring to the same real-world concepts (e.g., authors and inventors). The following steps are used:

\begin{itemize}
\item \textbf{Publication Citations:} Patent to Publications citations  in patents are  unstructured strings. We apply  a set of regular expression rules to extract titles, which can then be matched to OpenAlex \texttt{Work} nodes by name.

\item \textbf{Institution Matching:} Institutional names often differ across datasets (e.g., ``DOLBY LABORATORIES LICENSING CORPORATION'' vs. ``Dolby (United States)''). OpenAlex includes ROR (Research Organization Registry) IDs for institutions. We extract the ROR dataset,  normalize the  institution names (e.g., removing stopwords like ``Inc.'' or ``Corporation'') for ROR and  USPTO datasets, compute string similarity using Levenshtein distance to assign ROR IDs to USPTO institutions. Then the institutions from the two datasets can be joined by the ROR id. 

\item \textbf{Classification Alignment:} The two datasets have different hierarchal classification mechanism: OpenAlex uses ASJC (All Science Journal Classification), while USPTO uses CPC (Cooperative Patent Classification). We align CPC main groups with ASJC topics, CPC subclasses with ASJC subfields, and CPC classes with ASJC fields by encoding classification descriptions using sentence transformers and matching via vector similarity.
\end{itemize}
\noindent After  entity matching, entities can be joined by matching node properties with relational columns as shown in Table~\ref{tab:entity-matching}.

\begin{small}
\begin{table}[tb]
	\caption{Summary of OpenAlex node types and properties}
\label{tab:openalexnode}
\setlength{\tabcolsep}{1pt}
\centering
\begin{tabular}{|l|l|l|}
\hline
\textbf{Label} & \textbf{Node Properties} & \textbf{\#Nodes} \\
\hline
Author & \texttt{id},  \texttt{name}, \texttt{works\_count}, \texttt{cited\_by\_count} & 3,294,820 \\
Work & \texttt{id}, \texttt{name}, \texttt{year}, \texttt{type}, \texttt{cited\_by\_count} & 8,969,495 \\
Institution & \texttt{id}, \texttt{name}, \texttt{ror}, \texttt{works\_count}, \texttt{cited\_by\_count} & 112,271 \\
Field & \texttt{id}, \texttt{name} & 10 \\
Subfield & \texttt{id}, \texttt{name} & 89 \\
Topic & \texttt{id}, \texttt{name}, \texttt{works\_count}, \texttt{cited\_by\_count} & 1,571 \\
Keyword & \texttt{name} & 11,400 \\
\hline
\end{tabular}
\end{table}
\end{small}

\begin{small}
\begin{table}[tb]
\caption{Summary of some OpenAlex relationship types }
\label{tab:openalexedge}
\centering
\begin{tabular}{|l|l|l|r|}
\hline
\textbf{Relationship} & \textbf{Source Node} & \textbf{Target Node} & \textbf{\#Relationships} \\
\hline
WORK\_AT     & Author      & Institution  & 30,573,987 \\
CHILD\_OF    & Institution & Institution  & 22,060 \\
CREATED\_BY  & Work        & Author       & 16,244,003 \\
CREATED\_BY  & Work        & Institution  & 13,472,802 \\
RELATED\_TO  & Work        & Work         & 15,800,754 \\
ABOUT        & Work        & Topic        & 22,171,963 \\
HAS\_SIBLING & Topic       & Topic        & 67,510 \\
BELONGS\_TO  & Keyword     & Topic        & 15,705 \\
\hline
\end{tabular}
\end{table}
\end{small}

\begin{small}
\begin{table}[tb]\caption{USPTO  relational tables and schemas}\label{tab:usptodata}
\centering
\setlength{\tabcolsep}{1pt}
\begin{tabular}{|l|l|l|}
\hline
\textbf{Relation} & \textbf{Schema} & \textbf{\#Rows} \\
\hline
patent & \texttt{id}, \texttt{year}, \texttt{country}, \texttt{issue\_date}, \texttt{title} & 124,110 \\
pub\_cited & \textbf{\texttt{publication\_name}}, \texttt{patent\_ids} & 699,104 \\
inventors & \texttt{name}, \texttt{patent\_ids} & 210,573 \\
institution & \texttt{name}, \textbf{\texttt{ror}}, \texttt{patent\_ids} & 5,460 \\
maingroup & \textbf{\texttt{astj\_topic\_id}}, \texttt{symbol}, \texttt{patent\_ids}, $\cdots$& 418 \\
subclass & \textbf{\texttt{astj\_subfield\_id}}, \texttt{symbol}, \texttt{patent\_ids}, $\cdots$& 82 \\
class & \textbf{\texttt{astj\_field\_id}}, \texttt{symbol}, \texttt{patent\_ids}, $\cdots$& 16 \\
\hline
\end{tabular}
\end{table}
\end{small}

\begin{small}
\begin{table}[tb]
	\caption{Entity matching for nodes and relational columns}
\label{tab:entity-matching}
\centering
\begin{tabular}{|l|l|l|l|}
\hline
\textbf{Node Label. Property} & \textbf{Relation. Column} \\
\hline
Author.\texttt{name}     & inventors.\texttt{name} \\
Work.\texttt{name}      & pub\_cited.\texttt{publication\_name} \\
Institution.\texttt{ror}   & institution.\texttt{ror} \\
Field.\texttt{id}      & class.\texttt{astj\_field\_id} \\
Subfield.\texttt{id}     & subclass.\texttt{astj\_subfield\_id} \\
Topic.\texttt{id}     & maingroup.\texttt{astj\_topic\_id} \\
\hline
\end{tabular}
\end{table}
\end{small}

\subsubsection{Workloads}
We generate 500 cross-model queries using the LLM-based query generation method proposed in~\cite{zheng2024generating}. Among these, 465 queries are validated to be both syntactically and semantically correct. To simulate real-world filtering conditions, we inject random predicates on graph node properties and relational table columns. We randomly split the validated query set into 372 training  and 93 testing  queries. The test set is used to evaluate and compare  \textsc{MICRO} against baseline optimizers. The following example illustrates a benchmark query, where \texttt{neo4j} refers to the materialized result of a Cypher query:
\begin{cmls}{One benchmark query}
MATCH (a1:Author)<-[:CREATED_BY]-(w1:Work)-[:RELATED_TO]
   ->(w2:Work)-[:CREATED_BY]->(a2:Author)
WHERE a1.works_count > 50 AND a2.works_count > 50
RETURN w2.name AS w2name, w1.name AS w1name, a1.name AS a1name, a2.name AS a2name;
SELECT g.a2name, g.a1name, g.w2name, g.w1name
FROM neo4j g,publication_cited w,inventors a0,inventors a1,
WHERE g.a1name = a0.name AND g.a2name = a1.name
  AND g.w1name = w.name AND cardinality(a1.patent_ids) > 2;
\end{cmls}



\subsection{CM-LDBC Benchmark}
The LDBC Social Network Benchmark (SNB) is a widely-used synthetic graph benchmark for evaluating graph database performance. It models realistic social networks with entities such as people, posts, forums, tags and  organizations. Following the method in~\cite{zheng2024generating}, we adapt SNB to construct a semi-synthetic cross-model benchmark. Structural connections and node ids remain in the graph, while node attributes—including \texttt{id}—are stored in separate relational tables.

\begin{table*}[tb]
\centering
\caption{Statistics of CM-LDBC datasets (It shows $N_l$ for graph nodes and $S_l, TR_l$ for relations)}
\label{tab:cm-ldbc}
\renewcommand{\arraystretch}{1.2}
\begin{tabular}{lrrrrrrrrrrrrr}
\toprule
 & \textbf{Tag} & \textbf{TagClass} & \textbf{Comment} & \textbf{Forum} & \textbf{Person} & \textbf{Post}  & \textbf{Country} & \textbf{City} & \textbf{Company} & \textbf{University} \\
\midrule
SF-1    & 16,080  & 71  & 1,739,438  & 100,827  & 10,295  & 1,121,226    & 111  & 1,343  & 1,575  & 6,380 \\
SF-10   & 16,080  & 71  & 18,196,074 & 667,545  & 68,673  & 8,273,491   & 111  & 1,343  & 1,575  & 6,380 \\
\midrule
 T1     & 50K, 0.5 & 1K, 0.8 & 1M, 0.001 & 500K, 0.05 & 50K, 0.3 & 1M, 0.001 &  5K, 0.8 & 5K, 0.8 & 10K, 0.8 & 10K, 0.8 \\
 T2     & 100K, 0.1 & 1K, 0.8 & 1M, 0.0001 & 100K, 0.1 & 50K, 0.3 & 1M, 0.0001 & 5K, 0.5 & 5K, 0.5 & 50K, 1 & 50K, 1 \\
\bottomrule
\end{tabular}
\end{table*}

\subsubsection{Dataset}
The SNB datasets offer multiple scale factors; we use two: \texttt{SF-1} and \texttt{SF-10}. For each, we generate two sets of relational tables, $T1$ and $T2$, with varying sizes and join cardinalities to the node entities, resulting in four Cross-Model LDBC (CM-LDBC) datasets. For each node label $l$, we create a relational table containing all node properties. Let $N_l$ be the number of nodes with label $l$. We define the table size $S_l$ and true match ratio $TR_l$: we sample $N_l \times TR_l$ rows from the original node data as true join entries and synthesize the remaining $S_l - N_l \times TR_l$ rows with random attributes and synthetic \texttt{id}s not present in the graph. As a result, a cross-model join between graph nodes with label $l$ and the corresponding table yields exactly $N_l \times TR_l$ matches. Table~\ref{tab:cm-ldbc} summarizes these statistics: the first two rows report node counts; the last two rows report $S_l$ and $TR_l$ values.


These datasets are used for ablation study analyzing \textsc{MICRO}’s effectiveness under varied  conditions. Differences in table sizes affect data movement cost, while varying $TR$ affect the benefit of executing joins in Neo4j vs. PostgreSQL. For example, small and highly selective tables can improve performance when pushed into the graph by reducing intermediate Cypher result size with negligible table movement cost, while large but highly selective tables require trade-offs as moving them to graph database may incur significant overhead.

\subsubsection{Workloads}
We generate over 4,000 cross-model queries on CM-LDBC using the approach from~\cite{zheng2024generating}. As that work does not define a diversity metric, we perform manual filtering to ensure variation in Cypher paths structure and SQL join patterns. We select 505 distinct queries, split into 385 for training and 120 for evaluation to enable representative and fair comparisons across optimizers.

\section{Experiment}\label{sec:experiment}
We evaluate the performance of \textsc{MICRO} on the two proposed benchmarks and compare it against a baseline federated relational system, XDB~\cite{gavriilidis2023situ}. We also conduct ablation studies to assess the effectiveness of our learned optimizer, CMLero.

\subsection{Configuration}
For the OpenAlex-USPTO benchmark, the datasets are curated and managed by UCSD researchers. The OpenAlex graph is stored in a Neo4j instance, and the USPTO tables reside in a PostgreSQL database, each hosted on separate physical servers. \textsc{MICRO} is deployed on a third server provisioned via CloudLab~\cite{duplyakin2019design}, a public research infrastructure. The machine features a 24-core AMD 7402P CPU, 128 GB RAM, 1.6 TB of disk space, and runs Ubuntu 18.04.

For the CM-LDBC benchmark, the entire setup is deployed on a single CloudLab machine. We use Docker to host three containers—PostgreSQL storing relations, Neo4j storing graph data, and Python running \textsc{MICRO}—all connected within the same Docker network.

\subsection{Single-data model Setting}
We compare \textsc{MICRO} with XDB~\cite{gavriilidis2023situ}, a single-model federated relational system that, like MICRO, supports in-situ data access without data movement overhead across mediator and  databases. To support this comparison, we transform our cross-model benchmarks into a single-model format compatible with XDB. The relational part of the dataset remains unchanged. We made modifications to the graph data: 
\begin{itemize}
    \item Graph is transformed into relational tables: each node label or edge type becomes a separate table, with each row encode a node or edge and columns denote nodes/edges properties.
    \item These tables are loaded into a PostgreSQL server on the same physical node that previously hosted the Neo4j server.
    \item For index we built on previous graph node properties, we build similar index on the corresponding relational columns. 
\end{itemize}
We implement a simplified yet effective version of XDB: since there are only two relational servers, we designate the one with more data as the primary and use PostgreSQL's SQL/MED extension for accessing remote tables on the other Postgres server. Optimization features such as predicate and join pushdown are enabled through SQL/MED.

\subsection{Optimizers Baselines}
We compare  CMLero against three optimizer baselines: two rule-based baselines and one learned-model based: 

\begin{itemize}
    \item \textbf{Baseline-TS (Table Size):}  A threshold-based heuristic using table row count. For each query, if a table that joins with the graph has a row count smaller than a predefined threshold $t$, it is pushed into Neo4j for join execution.
    
    \item \textbf{Baseline-FVN (First-Visited Nodes):} Uses Neo4j's estimated  plan for the raw Cypher query to identify variables accessed by \texttt{NodeByLabelScan}. Tables joining with these variables are moved to Neo4j. Since such nodes are often  traversal entry points, early joins  may help filter them at the start of traversing, reducing  traversal cost.


    \item \textbf{Baseline-RLM (Regression Learned Model):} Shares CMLero’s architecture but omits the pairwise comparison layer. 
    During training, each plan is passed through the embedding layers, and its latency is used as the label. At inference time, the plan with the lowest predicted latency is selected.
\end{itemize}

\begin{figure*}[th!]
    \centering
    \includegraphics[width=0.98\textwidth]{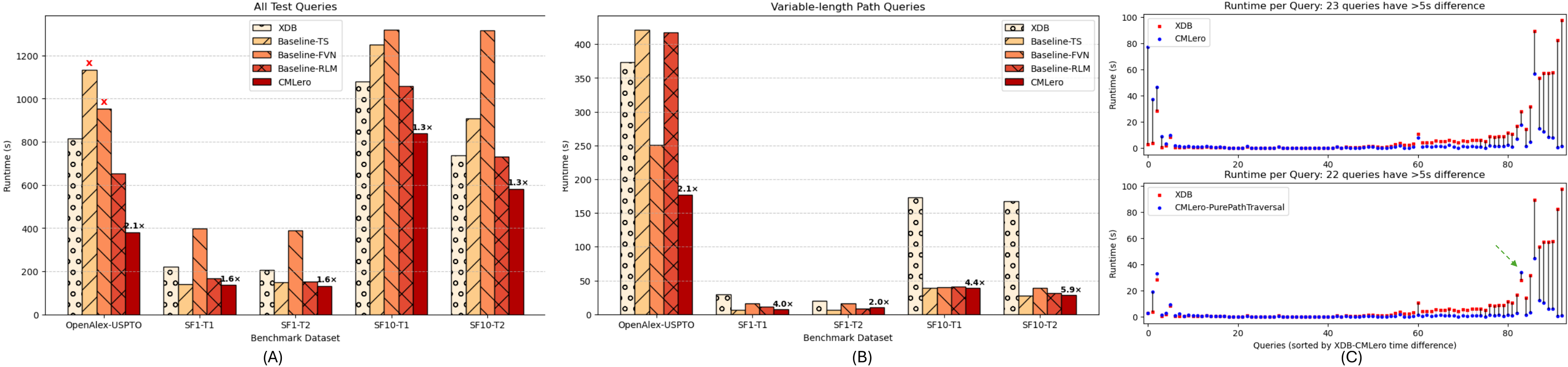}
    \caption{(A) Total runtime of all test queries across two benchmarks and five methods; (B) Total runtime of variable-length queries across two benchmarks and five methods; \change{(C) Per-query runtime for XDB and CMLero and  highlighted queries with time difference larger than 5 secs.}}\label{fig:exp-result}
\end{figure*}   
\subsection{End-to-End Results}
All results are reported on the test query set. Baseline-RLM and \textbf{CMLero} are trained on training queries.

Fig.~\ref{fig:exp-result} (A) shows the total runtime for all test queries across five datasets and five methods. A red mark indicate one query that exceeded the 5-minute timeout. Key observations include:

\textbf{Baseline-TS} exhibits varying performance across datasets: While it  outperforms XDB on some datasets (e.g., SF1-T1 and SF1-T2), it  performances worse on others. It also fails to complete one query within the timeout on  OpenAlex-USPTO. Besides, It requires manual tuning to determine an effective table size threshold. For each dataset, we experimented with thresholds from 1K to 500K rows and selected the value that yielded the best performance. The optimal threshold is 5K for the OpenAlex-USPTO benchmark and 100K for CM-LDBC. Notably, when database servers are deployed on different physical machines—as in OpenAlex-USPTO—the cost of moving large tables is significant, requiring smaller thresholds to avoid transferring large tables. 

\textbf{Baseline-FVN} consistently underperforms across all datasets. This is likely because the estimated plan for the raw Cypher query without alien tables does not match the actual execution plan after relational data is moved to Neo4j. Also, it only targets early filtering at path start nodes to reduce path traversal cost ignoring other factors such as  table movement cost and join selectivity between moved tables and nodes.

\textbf{Baseline-RLM} performs on par with or better than XDB across all datasets. \textbf{CMLero} consistently outperforms all baselines, achieving speedups of up to 2.1x, 1.6x, 1.6x, 1.3x, and 1.3x over XDB across all datasets. \change{The end-to-end runtime includes the total inference time, covering model loading, feature extraction, and evaluation for all test plans, which remains under 2 seconds even on CPU-only CloudLab nodes and is negligible compared to the total query execution time.}



Fig.~\ref{fig:exp-result} (C) upper part shows per-query runtime differences between CMLero and XDB on 93 test queries in OpenAlex-USPTO. Queries are sorted by time difference, with 23 queries exceeding a 5-second gap highlighted. CMLero is slower on 4 (4.3\%) and faster on 19 (20.4\%) of these. 


\subsection{Advantage of  Heterogeneous Database}
Different databases have distinct strengths. Graph engines like Neo4j are designed for efficient path traversal, which can be a bottleneck in relational systems. To evaluate this advantage, we isolate variable-length queries from  test workloads, 12 from OpenAlex-USPTO and  5 from CM-LDBC, and compare runtime performance across methods (Fig.~\ref{fig:exp-result} (B)).

The direct rewriting translates a variable length Cypher query to a  \texttt{WITH RECURSIVE} clauses defining the result table of recursive join, followed by  filters and joins with other tables. However, PostgreSQL handles such recursive queries inefficiently, particularly in the presence of dense relationships (e.g., \texttt{WORK\_RELATED\_TO\_WORK}), leading to significantly degraded performance. To ensure a fair comparison, we manually optimize the XDB implementation by pushing filters and joins into the base relations within the \texttt{WITH RECURSIVE} clause to improve execution efficiency. In contrast, \textsc{MICRO} requires no manual rewriting. Despite these efforts for XDB, \textsc{MICRO} achieves 2×–6× speedups on these path-intensive queries, demonstrating the clear advantage of leveraging native graph engines within a heterogeneous query execution framework.

\change{\subsection{Engine-specific Implementation of Rewrites}}
\change{We profiled the four queries where CMLero was slower than XDB and found that the bottleneck lay in executing the rewritten Cypher query. These queries involve node property joins between newly created nodes and original nodes combined with path traversal. For example, the second query joins topic and work entities across two databases. The plan chosen by CMLero (which was the best among candidates) produces the following Cypher query (we omit node property filter predicates and return clause for brevity):}

\begin{cmlsnc}
MATCH (w1:Work)-[:ABOUT]->(t:Topic)<-[:ABOUT]-(:Work)-
       [:CREATED_BY]->(:Institution)
MATCH (l:TopicT) MATCH (a:WorkT)  WHERE w1.name = a.name and t.id = l.id
\end{cmlsnc}

\change{Profiling this query in Neo4j shows that the engine performs the path traversal first, generating a large intermediate result, and only then applies the selective node property join. This yields around 4 million db hits in around 32 seconds, with the expensive value-hash join operator dominating cost. By contrast, XDB joins the small entity tables first, producing a much more efficient plan. Since Neo4j excels at path traversal, we replaced the node property join with explicit edges so that the query becomes a pure traversal. Concretely, when moving a table into Neo4j, we not only create nodes but also connect them to existing nodes if they share the same join property value. The node MATCH and WHERE clause then becomes:}

\begin{cmlsnc}
MATCH (t)-[:EQ_T]->(:TopicT) MATCH (w1)-[:EQ_W]->(:WorkT)
\end{cmlsnc}

\change{This version yields around 1 million db hits in 3.6 seconds  by pushing down the  path traversal denoting small entity join. We integrated this translator into MICRO, retrained it, and evaluated it on the test queries. Results are shown in the lower half of Fig.~\ref{fig:exp-result} (C), aligned with the same query order. The four slow queries improved substantially. However, not all queries benefited: e.g., the query marked by the green arrow became slower because data movement time increased dramatically when creating both nodes and edges for the moved tables. If data movement cost outweighs traversal savings, this approach is worse. This highlights that deeper optimization requires engine- and workload-specific knowledge, but in this paper we focus on a general implementation, so we keep our original implementation of the translator.}



\begin{figure}[htb]
\centering
    \centering
    \includegraphics[width=0.7\linewidth]{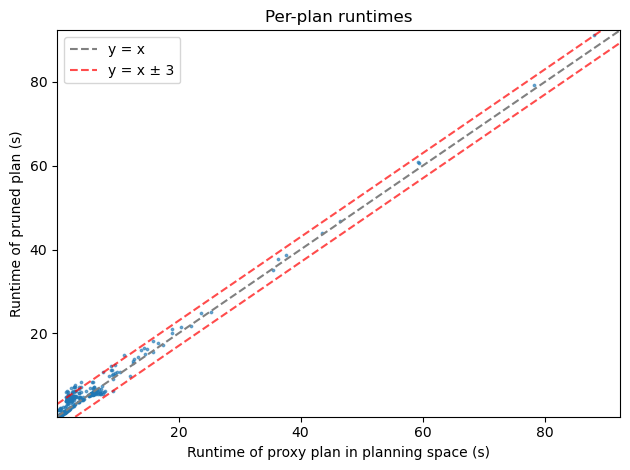}
    \caption{\change{Runtime comparison of pruned vs. retained plans.}}
    \label{fig:pruningruntime}
\end{figure}

\change{\subsection{Pruning Effectiveness and Training Cost}}
\change{To test the effectiveness of the pruning rule, for the test queries of OpenAlex-USPTO benchmark, we run the pruned plans which move vertices from Neo4j to Postgres, and move the result tables back to Neo4j (without moving other tables), and compare their runtime with the corresponding plans in our planning space which move the tables of the same entities to Neo4j instead. We compare their runtime in Fig.~\ref{fig:pruningruntime}, and out of the 634 plans, the retained plans perform better or similar in almost all cases; some pruned plans were faster  but less than three seconds.}



\begin{table}[htb]
\centering
\caption{\change{Training data collection stats.}}
\label{tab:prune}
\begin{tabular}{|l|l|l|}
\hline
                & OpenAlex-USPTO & SF10-T1 \\ \hline
\# plans        & 3192           & 4524    \\ \hline
\# pruned plans & 13977          & 21411   \\ \hline
collection time & 7 hr 15 min    & 15 hr 20 min  \\ \hline
\end{tabular}
\end{table}

\change{Table~\ref{tab:prune} shows that pruning avoids nearly four to five times as many plans. Given the small ($<3$s) advantage of the few better pruned plans, this loss is negligible relative to the training data collection time saved.}

\change{We further applied runtime-based pruning to stop long-running plans early. After materializing tables in Neo4j, we issue a \texttt{COUNT(<query>)} on the rewritten Cypher query. If the count takes longer than five minutes, we assign \texttt{EXTREME\_LARGE\_QUERY} as its runtime. If the count returns but exceeds a threshold where moving results back to Postgres would take $>5$ minutes, we assign \texttt{LARGE\_QUERY} time instead without running the  queries. With these rules, execution time for plans of training queries on OpenAlex-USPTO were collected in about 7h15m, and on the largest benchmark, SF10-T1, in about 15h20m.}

\subsection{Effectiveness of CMLero Optimizer}
We evaluate optimizer quality using two metrics:
\begin{itemize}
    \item \textbf{Q-Error}: The ratio of the latency of selected plan ($ST$) to the latency of the ground-truth best plan ($GT$): $Q = \frac{ST}{GT}$. We report the average Q-error across all test queries, along with the 90th and 95th percentile values.
    \item \textbf{Top-N Hit Rate (Top-N-HR)}: The proportion of test queries for which the selected plan falls within the top-N lowest latency plans: $\text{Top-N-HR} = \frac{|\text{Hit-Workloads}|}{|\text{Test-Workloads}|}$.
\end{itemize}

\noindent Table~\ref{tab:ablation-study} reports these metrics for all optimizers on the CM-LDBC benchmark. CMLero  consistently achieves the best performance across all datasets, significantly outperforming the other heuristic-based baselines and regression based baseline in both Q-error and Top-3 Hit Rate. Notably, CMLero not only reduces total runtime but also ensures that the selected plans for all queries are generally robust, as indicated  by the low 95th percentile Q-error.

\textbf{Baseline-TS} uses a fixed table size threshold across all CM-LDBC datasets, resulting in identical plans for SF1-T1 and SF10-T1 (and  for SF1-T2 and SF10-T2) when relation sizes are the same. However, its performance varies, indicating that table size alone is an insufficient criterion. It overlooks key graph-specific and cross-model factors—such as how a table's role may shift with changes in graph statistics—leading to suboptimal plans. While it performs well in certain cases, it is consistently outperformed by CMLero. \textbf{Baseline-FVN}, in contrast, uses graph-side information but neglects the relational side and how the moved table affects the graph query. As a result, it performs poorly across all datasets.

Between the two learned models, \textbf{Baseline-RLM}, despite sharing the same model architecture as CMLero, performs worse on all metrics. This is largely due to its regression-based objective, which struggles with limited training data. As a result, it often fails to capture subtle performance differences between plans and is less reliable in selecting optimal or near-optimal plans. For example, for the SF10-T1 benchmark, the training queries produce a total of 4499 valid plans leading to 4499 training samples for the \textbf{Baseline-RLM} model, however, 31461 pairs of plans are generated which provides \textbf{CMLero} much more training data.

\begin{table}[bt]
\centering
\caption{Optimizer comparison on CM-LDBC}
\label{tab:ablation-study}
\begin{tabular}{|l|l|c|c|c|c|}
\hline
\textbf{Data} & \textbf{System} & \textbf{Top-3-HR} & \textbf{AvgQ} & \textbf{Q90} & \textbf{Q95}  \\
\hline
SF1-T1  & Baseline-TS    & 0.66 & 7.67 & 3.54 & 34.93  \\
        & Baseline-FVN   & 0.20 & 18.96 & 10.07 & 24.04  \\
        & Baseline-RLM   & 0.67 & 5.13 & 2.74 & 3.54  \\
        & CMLero    & \textbf{0.82} & \textbf{1.19} & \textbf{1.52} & \textbf{1.65} \\
\hline
SF1-T2  & Baseline-TS    & 0.40 & 8.31 & 4.72 & 42.59  \\
        & Baseline-FVN   & 0.17 & 17.42 & 9.64 & 25.50  \\
        & Baseline-RLM   & 0.55 & 5.31 & 2.37 & 4.01  \\
        & CMLero    & \textbf{0.86} & \textbf{2.40} & \textbf{1.82} & \textbf{2.35} \\
\hline
SF10-T1 & Baseline-TS    & 0.68 & 7.14 & 3.12 & 6.57  \\
        & Baseline-FVN   & 0.32 & 5.27 & 8.12 & 15.31  \\
        & Baseline-RLM   & 0.51 & 5.65 & 3.34 & 4.90  \\
        & CMLero    & \textbf{0.78} & \textbf{2.50} & \textbf{1.68} & \textbf{2.48}  \\
\hline
SF10-T2 & Baseline-TS    & 0.68 & 8.20 & 2.96 & 7.87  \\
        & Baseline-FVN   & 0.23 & 5.88 & 8.55 & 14.71  \\
        & Baseline-RLM   & 0.63 & 3.82 & 3.54 & 5.48 \\
        & CMLero    & \textbf{0.83} & \textbf{1.99} & \textbf{1.53} & \textbf{2.55}  \\
\hline
\end{tabular}
\end{table}




\subsection{Adaptiveness to Distributed Environment}
In our initial CM-LDBC experiments, Neo4j and PostgreSQL were deployed as Docker containers on a single CloudLab machine, connected via a shared Docker network. To assess the robustness of \textsc{MICRO} in more realistic deployment scenarios, we extended this setup to a distributed environment across three physical machines. Specifically, we deployed Neo4j, PostgreSQL, and \textsc{MICRO} on separate CloudLab nodes, connected through a shared internal network.

\begin{figure}[bt!]
    \centering
    \includegraphics[width=0.47\textwidth]{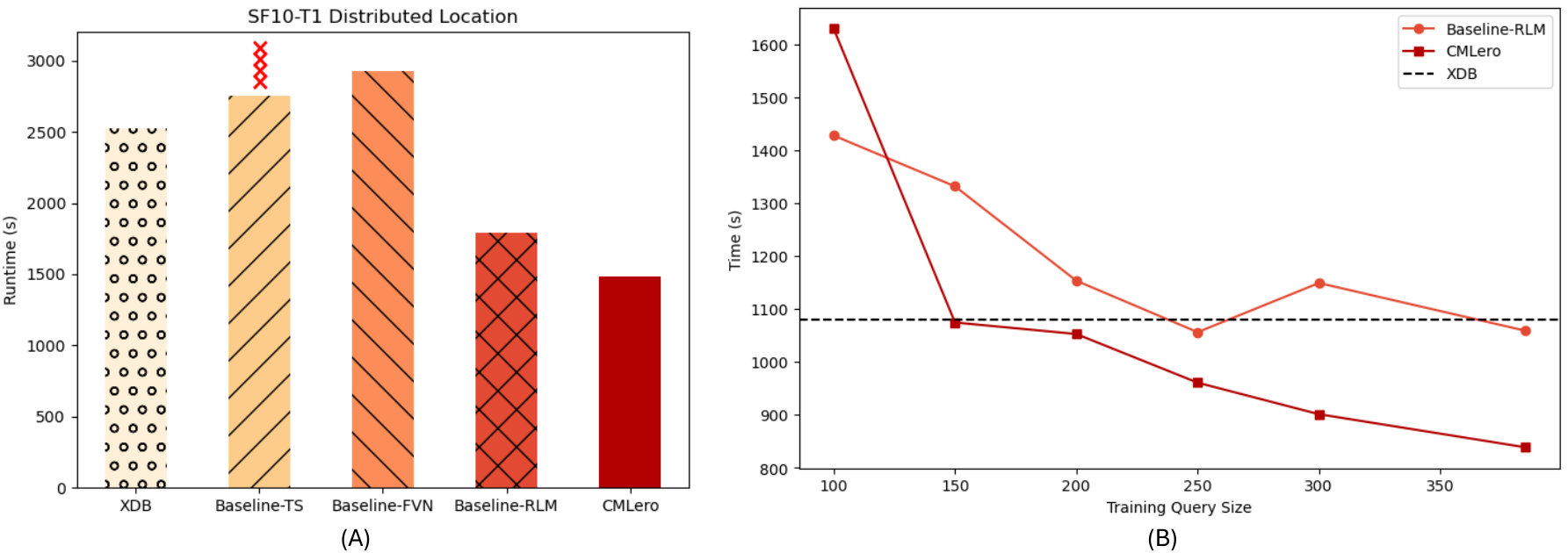}
    \caption{(A) Runtime of all test queries on SF10-T1 benchmark in distributed environment; (B)  Runtime of all test queries on SF10-T1 with different training sizes.}\label{fig:exp-geo}
\end{figure}

To emulate geographically distributed or cross-datacenter deployments, we imposed artificial network constraints between nodes: 50 Mbps bandwidth, 50 ms propagation latency, and a 64 KB burst allowance. We evaluate the SF10-T1 benchmark under this new setting with results  shown in Fig.~\ref{fig:exp-geo} (A). \textbf{Baseline-TS} has poor performance with 4 queries run out of time limit.  \textbf{CMLero} achieves around 1.7× speedup over XDB, a more substantial improvement than 1.3x in the single-node setup. In distributed settings, data movement between  databases  has higher cost, and by choosing the right tables to move to Neo4j, CMLero avoids more cost.  This  demonstrates its  adaptability  to various distributed environments.
 
\subsection{Optimizer Performance with Varying Training Query Sizes}
To evaluate the impact of training  size on the effectiveness of learned optimizers, for SF10-T1 benchmark, we vary the number of training queries from 100 to 385 and evaluate on the test query set. Fig.~\ref{fig:exp-geo} (B) presents the total runtime of test queries under both learned optimizers, compared to the XDB baseline. \textbf{CMLero} reaches performance comparable to XDB with only 150 training queries and continues to improve as the training size increases. \textbf{Baseline-RLM}, however,  requires around 250 queries to match XDB-level performance, and  shows limited improvement beyond that point, indicating lower sample efficiency and less effective generalization.

\subsection{\change{Limitations and Discussion for Future Work}}
\change{For ease of benchmark generation and model design, we currently restrict focus to a limited query pattern. More complex cases—such as nested joins or multi-round bidirectional joins—can arise in practice. An  extension of CMLero is to decompose such queries into multiple Cypher+SQL segments, apply the model to each segment, and aggregate their scores into a final prediction. Future work may also explore attention mechanisms across segments to better capture interactions. This extension does not require a redesign of the overall architecture, but rather incremental additions.}

\change{Besides, our experiments use offline training on a fixed benchmark, however, the framework can be extended when the data updates or fresh: the learned model can be incrementally retrained starting from existing parameters rather than from scratch, reducing training cost and preserving prior knowledge.  CMLero uses  estimated plan from Neo4j collected statically offline, adaptive re-optimization maybe needed when the actual Neo4j  plan deviates a lot from the estimates. Since  all candidate plans for a query share the  Neo4j's  plan for the original Cypher query, recollecting the real query plan can benefit the evaluation of  all candidates. Exploring continuous or online fine-tuning is an interesting direction for future work.}



\section{Related work} \label{sec:related}

\underline{\textbf{ML-Based Query Optimization for RDBMS.}} ML-based methods have been developed for query  optimizers. A majority of them work on learned  cardinality estimations (CE) and fall into two groups: data-driven approaches that model data distributions using generative models \cite{DBLP:journals/pvldb/YangKLLDCS20,DBLP:journals/corr/abs-2012-14743,DBLP:journals/pvldb/HilprechtSKMKB20,DBLP:journals/pvldb/ZhuWHZPQZC21}, and query-driven models trained on sampled queries \cite{dutt2019selectivity,wu2021unified,DBLP:conf/cidr/KipfKRLBK19}. 
Experiment \cite{10.14778/3461535.3461552} shows that these methods significantly outperform traditional techniques at CE, and integrating them into DBMSs has yielded real performance gains \cite{10.14778/3503585.3503586,zhang2023autoce}. Some ML approaches focus on learned cost model \cite{sun2019end,zhi2021efficient,hilprecht2022zero} which utilize neural networks to estimate the cost of a query; systems like Neo \cite{10.14778/3342263.3342644}, Balsa \cite{10.1145/3514221.3517885} and Bao \cite{marcus2021bao} formulate  plan enumeration  as a  Reinforcement Learning (RL) problem and designs  deep neural networks such as TCNN as the value network  to predicate the latency of  plans to choose the optimal one. As compared in Section~\ref{sec:intro-lero}, these pointwise approaches have certain limitations and  Lero ~\cite{zhu2023lero} provides pairwise approach which overcomes these shortcomings.   

\underline{\textbf{Query Optimization in Federated Database System.}} Federated Database System (FDBS) integrates multiple  databases into a unified system. Queries are decomposed and delegated to different databases and their results are integrated. Optimization must account for engine-specific costs.  Early systems like Garlic and Disco~\cite{roth1997don,haas1997optimizing,tomasic1998scaling} use site-specific   wrappers to extract cost/cardinality estimates;  \cite{deshpande2002decoupled} sends bids to the underlying system to acquire the cost of an operation; \cite{du1992query,gardarin1996calibrating} calibrate cost formulas for each system via execution of a large set of queries. More recently, XDB~\cite{gavriilidis2023situ} adopts a mediator-free model leveraging SQL/MED for in-situ query execution. ML-based approaches also exist: \cite{xu2019learning} uses random forests to  choose a single engine as the federated engine per query; Coral~\cite{gu2023coral} optimizes join orders and engine selections  using a  RL model  with deep Q-networks  estimating the reward of each join action. However, these systems assume homogeneous  relational data model, leaving cross-model challenges unaddressed.

\underline{\textbf{Query Optimization in Polystore system.}}
Polystores is a specialized FDBS that integrates engines with different data models. BigDAWG \cite{elmore2015demonstration,duggan2015bigdawg,gadepally2016bigdawg} groups engines into model-specific ``islands'', optimizes mainly within single islands by profiling operations on different engines within the island, and requires manual CASTs for cross-island queries, thus avoiding cross-model join optimization. ESTOCADA \cite{alotaibi2020estocada} focuses on query rewriting leveraging materialized views in different models,  which focuses on a different optimization goal than ours. Wayang \cite{agrawal2016rheem,beedkar2023apache,kruse2020rheemix,agrawal2018rheem} supports a variety of platforms including DBMSs and Spark through fine-grained API where fine-granular keyword/operator can be  mapped to  different platforms. Optimization selects platforms with the lowest cost per operation, but this approach does not preserve DBMS autonomy or utilize native query optimizers.

\underline{\textbf{Cross-Model Join Optimization.}}
A few works explore cross-model joins. \cite{zhang2020selectivity} estimates selectivity for relational-tree joins. \cite{fu2024joining,jindal2015graph,zhao2017all,dave2016graphframes} supports relational-graph hybrid queries by extending relational model for graph, enabling unified query planning.  However, both assume a single RDBMS extended to support other data model—unlike our heterogeneous multi-engine setting.

\section{Conclusions and Future Work} \label{sec:con}
In this paper, we defined a class of cross graph-relational join queries as an important subset of cross-model cross-engine queries. We introduced a real-world and several semi-synthetic benchmarks to support research on query processing over heterogeneous data models. We presented MICRO, a lightweight middleware designed to efficiently execute such queries without incurring the overhead of an inflated plan  space. Despite its simplicity, experimental results demonstrate that MICRO achieves strong performance on complex workloads. At the core of MICRO, the learning-to-rank optimizer CMLero consistently outperforms both rule-based and regression-based baselines, underscoring its effectiveness in selecting efficient execution plans in cross-model settings.

This work serves as an initial step toward addressing the broader challenges of federated query optimization across heterogeneous data models. Currently, the system supports one graph engine and one relational engine with multiple tables. Future work includes extending the architecture to support
other cross-model operators, multiple graph and relational engines, and their interplay with indexes, all of which will require  generalizing the design of CMLero. Beyond graph and relational models, real-world applications often involve additional data modalities, such as unstructured text served by specialized systems like Solr. Enabling efficient query processing across a wider spectrum of data models remains an important direction for future research.

\section*{Acknowledgment}
This work was supported in part by NSF under award numbers 1942724 and 1909875. The content is solely the responsibility of the authors and does not necessarily represent the views of NSF.
\clearpage
\bibliographystyle{IEEEtran}
\bibliography{sample}
%


\end{document}